\documentclass[structabstract]{raa_twocloumn}
\usepackage{graphicx,times}
\usepackage{natbib}
\usepackage{verbatim}
\usepackage{booktabs}
\usepackage{amssymb,amsmath}
\usepackage{longtable}
\bibpunct{(}{)}{;}{a}{}{,}

\usepackage[a4paper=true,pagebackref=true]{hyperref}
\hypersetup{pdftitle = The title of my PDF, pdfauthor = My name, pdfsubject= The subject, pdfkeywords = keyword1 keyword2 keyword3} 
\hypersetup{colorlinks = true, linkcolor = green, anchorcolor = red, citecolor = blue, filecolor = red, pagecolor = red, urlcolor = red}

\newcommand{\Msun}           {\,{\rm M}_\odot}

\newcommand{\nii} {[{N~\sc ii}]$\lambda$6584}
\newcommand{\sii} {[{S~\sc ii}]$\lambda\lambda$6717,6731}

\begin{document}

   \title{Spatially Resolved Properties of Supernova Host Galaxies in SDSS-IV MaNGA}

 \volnopage{ {\bf 20XX} Vol.\ {\bf X} No. {\bf XX}, 000--000}
   \setcounter{page}{1}

   \author{Hong-xuan Zhang\inst{1,2,3}, Yan-mei Chen\inst{1,2,3}, Yong Shi\inst{1,2,3}, Min Bao\inst{1,2,3}, Xiao-ling Yu\inst{1,2,3}
   }

   \institute{ School of Astronomy and Space Science, Nanjing University, Nanjing 210093, China\\
   \and
                    Key Laboratory of Modern Astronomy and Astrophysics (Nanjing University), Ministry of Education, Nanjing 210093, China\\
                    \and
                    Collaborative Innovation Center of Modern Astronomy and Space Exploration, Nanjing 210093, China; {\it chenym@nju.edu.cn}\\
\vs \no
   {\small Received 20XX Month Day; accepted 20XX Month Day}
}

\abstract{We crossmatch galaxies from Mapping Nearby Galaxies at Apache Point Observatory with the Open Supernova Catalog, obtaining a total of 132 SNe within MaNGA bundle. These 132 SNe can be classified into 67 Type Ia and 65 Type CC. We study the global and local properties of supernova host galaxies statistically. Type Ia SNe are distributed in both star-forming galaxies and quiescent galaxies, while Type CC SNe are all distributed along the star-forming main sequence. As the stellar mass increases, the Type Ia/CC number ratio increases. We find: (1) there is no obvious difference in the interaction possibilities and environments between Type Ia SN hosts and a control sample of galaxies with similar stellar mass and SFR distributions, except that Type Ia SNe tend to appear in galaxies which are more bulge-dominated than their controls. For Type CC SNe, there is no difference between their hosts and the control galaxies in galaxy morphology, interaction possibilities as well as environments; (2) the SN locations have smaller velocity dispersion, lower metallicity, and younger stellar population than galaxy centers. This is a natural result of radius gradients of all these parameters. The SN location and the its symmetrical position relative to the galaxy center, as well as regions with similar effective radii have very similar [Mg/Fe], gas-phase metallicity, gas velocity dispersion and stellar population age.
}

   \authorrunning{Zhang H.~X. et al. }            
   \titlerunning{Properties of Supernova Host Galaxies}  
   \maketitle

%
\section{Introduction}           
Supernova (SN) explosion, as an important astrophysics process, can be classified into Type I and II according to the presence of hydrogen line or not in their spectra \citep{fil97}. Type I Supernovae (SNe) are divided into Type Ia with Si 6510\r{A} absorption line, Type Ib without Si but with He line, and Type Ic neither with Si or He line in the spectra \citep{ham02}. 
Type II SNe can be divided into Type IIP and Type IIL \citep{bar79} by light curves that Type IIP has a long plateau ($\sim$90days) after the curve reaches its peak \citep{rub16} while Type IIL shows a linear decline after the peak.

In another way, we can also classify SNe by their progenitors. The explosion of a progenitor star with stellar mass $>$ 8 $M_{\odot}$ is induced by the gravitational collapse and leaves a neutron star or black hole in the center \citep{arn89,bet79}. This is named as core collapse supernovae (CC SNe), which is corresponding to Type Ib/c and Type II SNe. Meanwhile, a star with stellar mass in between 0.5$\sim$8 $M_{\odot}$ will evolve into a degenerate carbon-oxygen (C/O) white dwarf (WD, \citealt{bec80}) as its end. If the mass of the WD is larger than $\sim$1.44 $M_{\odot}$, thermonuclear reactions can ignite the center of the WD and lead to a great thermonuclear explosion which is named Type Ia SNe \citep{hoy60}. 

As a result, Type Ia SNe have been recognized as an important cosmological probe. Because of their stable peak luminosity, it is treated as the best standard candle up to date. It provides the evidence of accelerating expansion of the universe and the existence of dark energy \citep{per99,rie98}. However, there have been several studies showing that the peak luminosity of Type Ia SNe varies with host galaxies properties, which would lead to the uncertainties in constraining cosmological models. At the early stage, these studies focused mainly on the global properties of host galaxies, and how the peak luminosity of Type Ia SNe depends on stellar masses \citep{sul10}, ages \citep{ham95,ham00} and gas metallicities \citep{gal05}.

As these kinds of investigation went deeper and deeper, astronomers realized that the local environments of SNe should be important in influencing the properties of SNe. However, there is no large spatial resolved surveys by that time, the studies are focused on the locations where the supernova explodes \citep{and09,galb12} since most galaxy properties have obvious radial gradients.

With the development of observation technology, the emergence of integral field unit (IFU) survey provides spatial resolved information of galaxies and makes it possible to construct a direct observation of local properties at SN exploding location. \cite{sta12} compared local and global properties and environments of host galaxies of 6 Ia SNe and found no obvious difference in metallicity based on the Calar Alto Legacy Integral Field Area (CALIFA) survey. \cite{galb14,galb16} expanded the sample size of SN host galaxies to 115. By the PMAS/PPAK Integral-field Supernova hosts Compilation (PISCO), \cite{galb18} has enlarged the sample to 232 host galaxies with 272 SNe.

\cite{zho19} compared properties of SN exploding locations with global properties of their host galaxies based on MaNGA data, including 4 Type Ia, 5 Type II and 2 unclassified SNe, and found that all of the SNe are exploded in metal-rich galaxies with $\rm{12+log(O/H)>8.5}$ but showed small difference between local and global metallicity. In 2020, \citeauthor{zho20} revisited the same question and found similar results with a larger sample of 67 SNe.

We notice that most of the previous researches based on IFS observations do not take the radial gradients into account. However, the comparison between local and global properties could be influenced by radial gradients. In this work, we build a sample of 132 SNe with MaNGA observations of host galaxies, comparing the properties of SN location with different regions in the host galaxies.

The paper is organized as follows: in Section 2, we describe our sample selection criteria. Then we show the comparison results in aspects of both global and local properties in Section 3. The results and conclusions are shown in Section 4 \& 5. The cosmological parameters we use throughout this paper are $H_{\rm0}$ = 70 km s$^{-1}$ Mpc$^{-1}$, $\Omega_{\rm M}$ = 0.3, and $\Omega_{\rm \Lambda}$ = 0.7.

\section{Observations and Data Reduction} 
\subsection{Mapping Nearby Galaxies at Apache Point Observatory}

MaNGA \citep{bun15}, as one of three major programs in Sloan Digital Sky Survey IV \citep{bla17}, use Apache Point Observatory's 2.5-meter telescope \citep{gun06} for observation. MaNGA launched in 2014 July, and was completed in early 2021, obtaining the largest spatial resolved galaxy sample so far ($\sim$10,000 galaxies). In order to observe the distribution of the gas, stellar kinematics and stellar populations in galaxies of different sizes, the MaNGA IFUs include different sizes of fiber-bundles. Each cartridge includes 2$\times$N$_{19}$, 4$\times$N$_{37}$, 4$\times$N$_{61}$, 2$\times$N$_{91}$, 5$\times$N$_{127}$, altogether 17 different sizes of bundles as well as 12$\times$N$_{7}$ mini-bundle per cartridge for flux calibration \citep{dro15,yan16a}. The two dual-channel BOSS spectrographs \citep{sm13} provide a simultaneous wavelength range of 3600$\sim$10300$\rm \AA$ with a spectral resolution of R$\sim$2000. As a result of 3 hours in typical integration time, the $r-$band signal-to-noise ratio (S/N) reaches 4$\sim$8 at 23 mag arcsec$^{-2}$.

The target sample covers a range in redshift 0.01 $\sim$ 0.15 with a median redshift of 0.03. The stellar mass has a flat distribution in between $10^{9} M_{\odot}$ to $10^{11} M_{\odot}$ for ensuring adequate sampling of stellar mass. The MaNGA galaxies are separated into two different subsamples: the primary sample includes more than 80$\%$ of all targets and covers to 1.5 effective radii (R$_e$) and the secondary sample to 2.5R$_e$.

The MaNGA data used in this work include observations of 8000 unique galaxies in MaNGA Product Launch-9 (MPL9). The raw data is reduced by Data Reduction Pipeline (DRP,  \citealt{law16}) and then analyzed through the Data Analysis Pipeline (DAP, \citealt{wes19,bel19}).

\subsection{The Open Supernova Catalog}

The Open Supernova Catalog (TOSC, \citealt{gui17}) is a website dedicated to collecting supernova data\footnote{\url{https://sne.space}}. It has collected nearly 78,000 supernova and candidates till March 2021, most of which come from public supernova catalogs and projects, such as Asiago Supernova Catalog (ASC, \citealt{bar99}), Caltech Core-Collapse program (CCCP, \citealt{gal04}), Cambridge Photometry Calibration Server (CPCS, \citealt{zie19}), etc. It also collects individual supernovae in published papers (e.g., \citealt{yar12,mat08,sil12}). This website provides basic parameters including SN name, type, position, redshift, and whether it contains X-ray and radio observations. Currently, about 20,000 supernovae in the catalog contain light curves, and about 10,000 contain spectra. This website has laid a solid foundation for the systematic researches of supernovae. 

\subsection{Sample Selection and Data Analysis}\label{2.3}

At first, we crossmatch 8000 galaxies from MPL9 with 77951 SNe from TOSC with a tolerance radius of 3R$_e$ of the MaNGA galaxy and obtain 411 SN host galaxies. Meanwhile, there is a MaNGA ancillary program planned to observe host galaxies of 60 type Ia supernova (Accetta et al. 2021, submitted). In MPL9, 7 of these 60 targets have been observed and we find that three of them are included in our sample. We then exclude 188 SNe as well as their host galaxies from our sample since the type of these SNe is not clear. we also remove 47 SNe which are located outside the MaNGA bundles. The stellar mass and SFR are taken from MPA-JHU\footnote{\url{https://wwwmpa.mpa-garching.mpg.de/SDSS/DR7/}} catalog. The total stellar mass is estimated through the fitting of SDSS $ugriz$ 5-band SED \citep{kau03}. The SFR is estimated from the luminosity of the H$\alpha$ emission line \citep{bri04}. Finally we get a sample of 132 SNe, which includes 67 Type Ia and 65 Type CC SNe (8 Type Ib, 6 Type Ic and 51 Type II). The basic information of these SNe is listed in Table \ref{table:sne}. 

MaNGA DAP \citep{wes19} fits the galaxy's spectrum based on MaStar Library \citep{yan19} using pPXF spectrum processing software \citep{cap04}. The DAP fits the stellar continuum and 21 nebular emission lines of main optical bands for each spaxel. The parameters extracted in this work are from the DAP named ``SPX-MILESHC-MASTARHC" including nebular emission line fluxes (\nii\ \& \sii\/), the spectral indices ([Mg/Fe] \& D$n$4000) and velocity dispersion of ionized gas ($ \sigma_{{\rm H\alpha}}$), etc.

\begin{figure}
\centering
\includegraphics[width=0.5\textwidth]{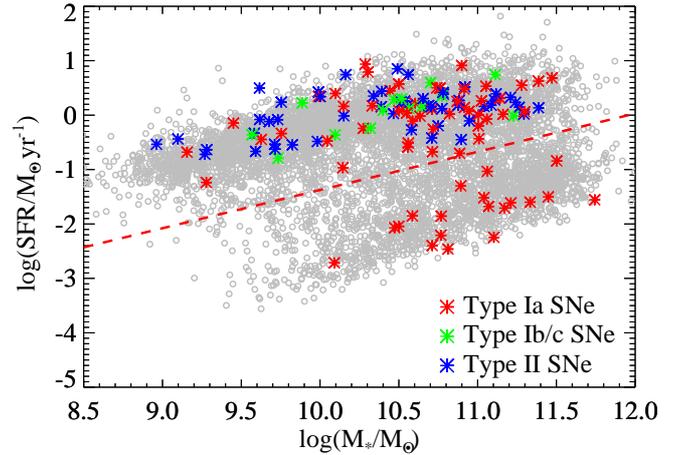}
\caption{The SFR vs. $M_*$ diagram of Type Ia (red asterisk), Type Ib/c (green asterisk) and Type II SNe (blue asterisk), the grey circles in background are from MaNGA MPL9 and the red dash line is 2$\sigma$ below the star forming main sequence.}
\label{contour1}
\end{figure}

\begin{figure}
\centering
\includegraphics[width=0.5\textwidth]{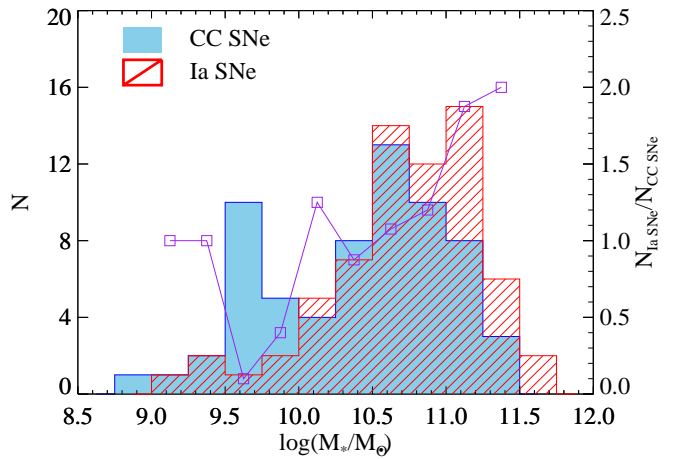}
\caption{The stellar mass distribution of Type Ia (red) and Type CC (blue) SN host galaxies. The purple open squares and solid line represent the Ia/CC SNe number ratio in each bin.}
\label{n_cc}
\end{figure}

\begin{figure}
  \centering
  \includegraphics[width= 0.45\textwidth]{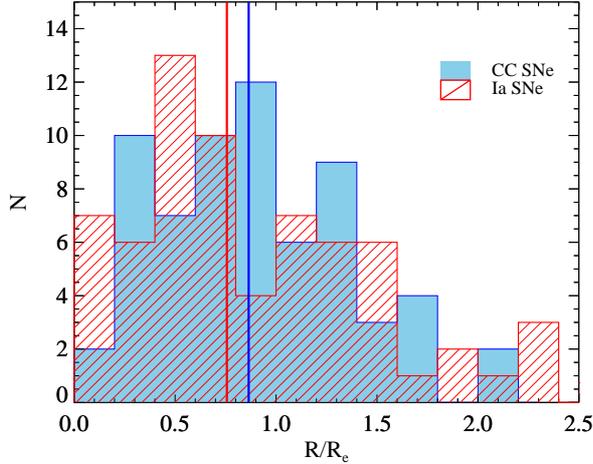}
  \caption{The distribution of the SN position relative to the galaxy center  in unit of R$_e$. Red histogram for Type Ia and blue for Type CC SNe. Red and blue solid lines represent the median values.}\label{eff}
\end{figure}

Figure \ref{contour1} shows the distributions of Type Ia (red asterisk), Type Ib/c (green) and Type II (blue) SN host galaxies in SFR vs. $M_*$ diagram. The grey circles represent MaNGA MPL9 galaxy sample, and the red dash line separates the star-forming main sequence with higher SFRs from quiescent galaxies. We can find that Type CC SNe (including Type Ib/c and Type II) are all located in the star-forming main sequence above the dash line, which corresponds to the fact that the progenitor of these SNe are all massive young stars and formed in star-forming region. Meanwhile, Type Ia SNe distribute in both blue and red galaxies because the progenitors of Ia SNe are all low-mass stars, which exists in both star-forming and quiescent galaxies. This result is consistent with \cite{sha14} that among their 902 SNe, Type CC SNe are most distributed in star forming galaxies and a small part in AGN and quiescent galaxies while Type Ia distributed in both types of galaxies.

Figure \ref{n_cc} shows the stellar mass distribution of Type Ia (red histogram) and Type CC (blue) SN host galaxies. The purple squares and line shows the ratio between the number of Type Ia and CC SNe. At log$M_*$/$\Msun$ $<$ 9.5, the statistical uncertainty of number ratio is large due to that there are only 3 Type Ia and 4 CC SNe in this mass range. At log$M_*$/$\Msun$ $>$ 9.5, the ratio increases quickly with stellar mass, consistent with the fact that the median stellar mass of Ia host galaxies is higher than Type CC host galaxies in \citet{zho20}. 

In Figure \ref{eff}, we show the distribution of the SN position relative to the galaxy center  in unit of R$_e$. The red histogram represents Type Ia SNe, blue represents CC and the red and blue vertical lines represent the median values of Ia and CC SNe, respectively. The distribution of the two SN types shows no significant difference. The sharp decrease of SN number outside 1.5R$_e$ is due to the fact that only 20$\%$ of MaNGA bundles cover beyond 1.5R$_e$.

\begin{figure*}
\centering
\includegraphics[width=0.8\textwidth]{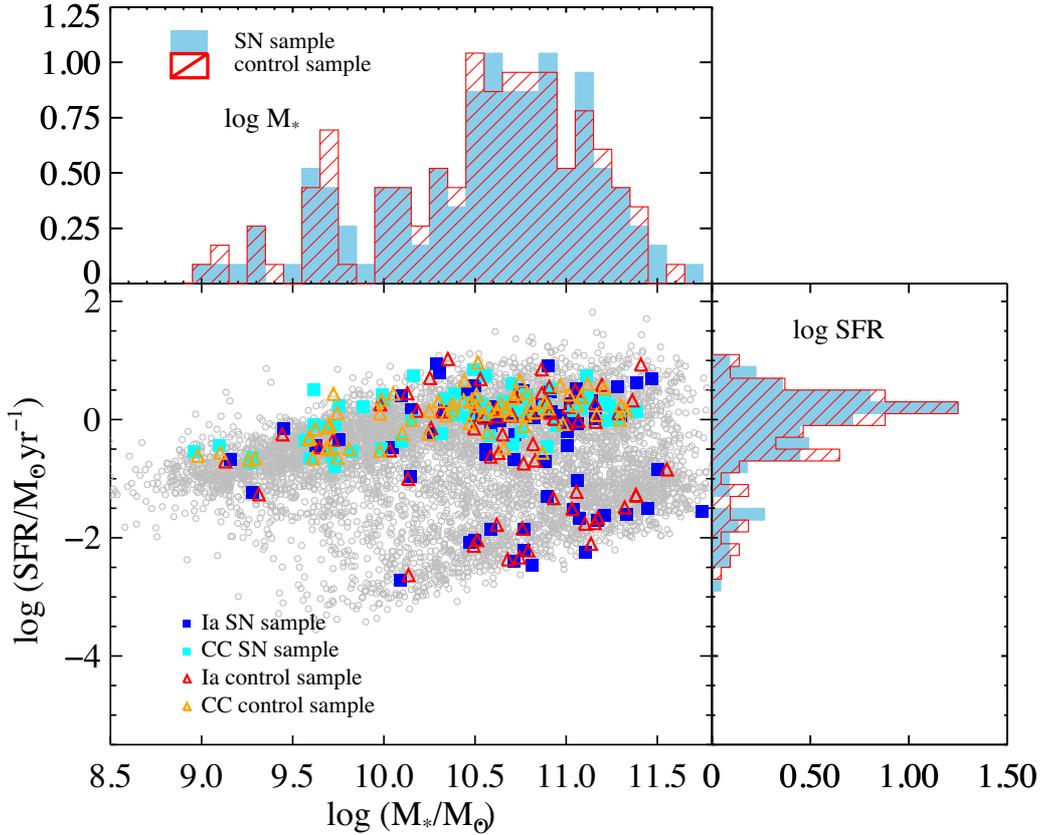}
\caption{The SFR vs. $M_*$ diagram of SN sample (blue and cyan solid square for Type Ia and Type CC sample) and control sample (red and orange open triangle for Type Ia and Type CC sample) and MaNGA MPL9 sample (grey open circles), the top and right panels show histograms of $M_*$ and SFR, respectively. In each panel, the histogram filled with blue color is for the SN sample and that filled with red strip lines is for the control sample.}
\label{con2}
\end{figure*}

In order to compare the properties of SN host galaxies and non-SN host galaxies, we build a non-SN host galaxy control sample, which is closely matched in $M_*$, SFR and $z$. The specific selection method is shown in Equation \ref{equ:2}.

\begin{equation}\label{equ:2}
\centering
\Delta s =
|\frac{\Delta {\rm log }M_*}{0.1}|+|\frac{\Delta {\rm log SFR}}{0.2}|+|\frac{\Delta z}{0.001}|
\end{equation}
We choose galaxy with smallest $\Delta s$ as SN control where $\Delta {\rm log}  M_*$ represents the stellar mass difference between the SN host galaxy and its control, $\Delta {\rm log  SFR}$ and $\Delta z$ represent the SFR and redshift difference, respectively. The motivation of choosing these three parameters to build control sample is the following: (i) as one of the fundamental parameters, stellar mass correlates strongly with many other parameters of galaxies such as metallicity and SFR; (ii) the SFR is a major factor of SN explosions especially for Type CC SNe, a high SFR represents a high SN explosion rate; (iii) a similar redshift distribution is required to avoid the influence of any evolutionary effects. Figure \ref{con2} shows the distributions of SN host galaxies and the control sample in SFR vs. $M_*$ diagram. The blue and cyan solid squares correspond to Type Ia and Type CC SN host galaxies. The red and orange open triangles correspond to the control galaxies of Type Ia and Type CC, respectively. The top and right panel show the histograms of $M_*$ and SFR for the SN sample (blue) and control sample (red). It is clear that the SN sample and the control sample have similar $M_*$ and SFR distributions.

\section{Results}

In this section, we compare the morphology, interaction possibility parameter and environment of the SN host galaxies and control sample, trying to understand the correlation between these properties and SN explosions. Thanks to the large IFU sample provided by MaNGA survey, we can compare the properties of the local region of SN explosions and other locations within the galaxy, investigating the connection between the SN explosions and the properties of their host galaxies.

\subsection{The Global Properties of SN Host Galaxies}

We use S{\`e}rsic index $n$ to characterize the morphology of galaxies \citep{ser63,ser68}. The surface brightness of a galaxy can be fitted by S{\`e}rsic profile:

\begin{equation}\label{sersic}
\centering
I(R)=I_0exp[-\beta_n(\frac{R}{R_e})^{1/n}]
\end{equation}
where $I_0$ is the central surface brightness of the galaxy and the parameter $n$ in Equation \ref{sersic} represents the S{\`e}rsic index. $n$ = 2 is generally used as the threshold between disk galaxies ($n<$ 2) and elliptical galaxies ($n>$ 2). We take $n$ of our sample from PyMorph\footnote{\url{https://www.sdss.org/dr15/data_access/value-added-catalogs/?vac_id=manga-pymorph-dr15-photometric-catalog}} \citep{fis19} in MaNGA data release 15 (DR15). MaNGA Morphology Deep Learning\footnote{\url{https://www.sdss.org/dr15/data_access/value-added-catalogs/?vac_id=manga-morphology-deep-learning-dr15-catalog}} \citep{dom18} provides a parameter named $P_{merg}$ for DR15, which represents the likelihood that the galaxy has recently experienced interactions by optical images \citep{wil13}. The environment parameters $d_{1st}$ and $N_{neigh}$ are taken from GEMA-VAC\footnote{\url{https://www.sdss.org/dr15/data_access/value-added-catalogs/?vac_id=gema-vac-galaxy-environment-for-manga-value-added-catalog}} \citep{god17}, where $d_{1st}$ is the projection distance between the target galaxy and the nearest neighbour with $r$-band magnitude $m_r$ $<$ 17.7. $N_{neigh}$ is the total number of neighbor galaxies with $m_r$ $<$ 17.7, a line-of-sight velocity to the target galaxy within $\pm$500km s$^{-1}$, the projection distance within 1Mpc.

\begin{figure*}
  \centering
  \includegraphics[width=0.9\textwidth]{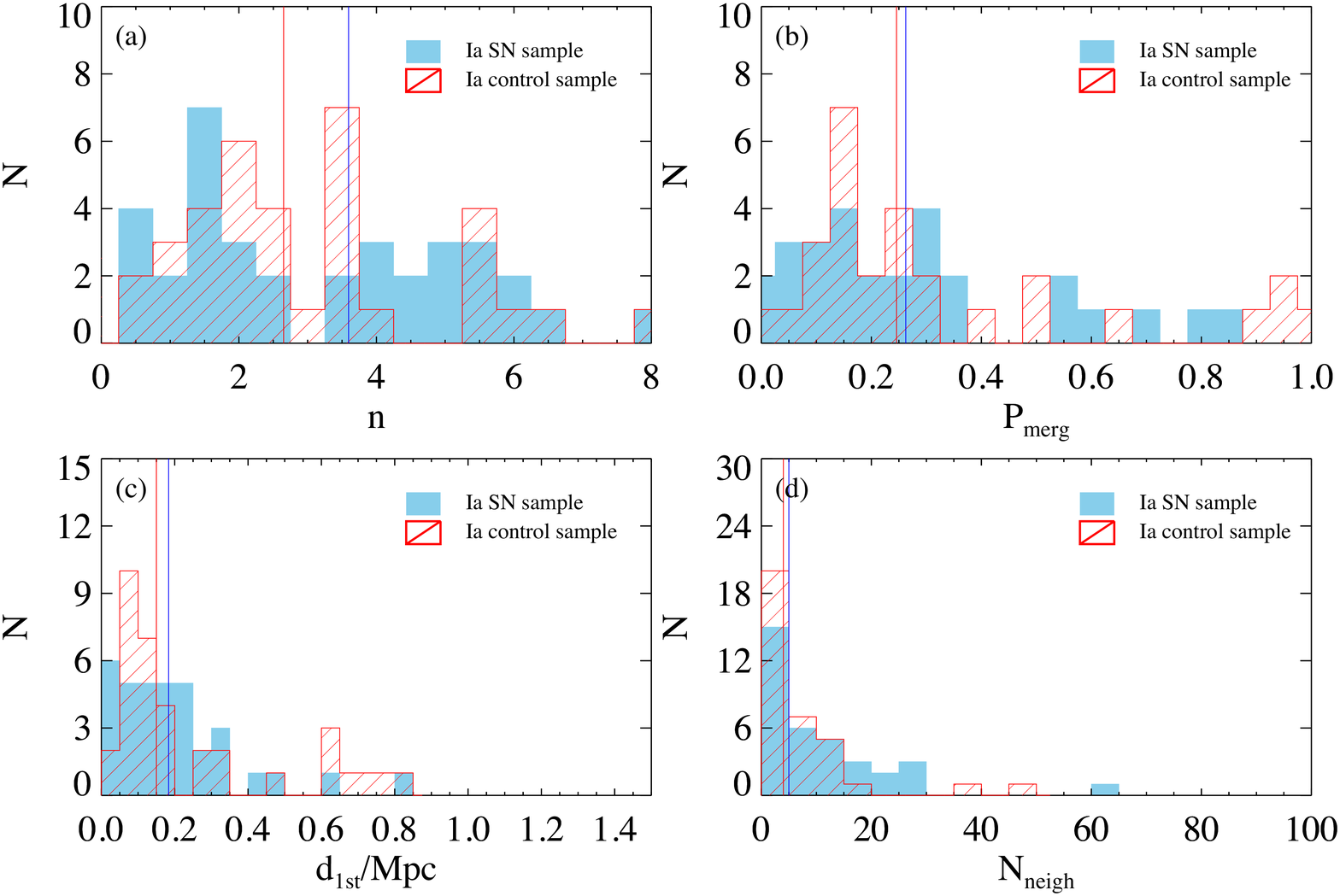}
  \caption{The distribution of four parameters of Type Ia SN host galaxies (blue) and their control sample (red).  {\bf (a)} S{\`e}rsic index $n$;  {\bf (b)} merging possibility $P_{merg}$;  {\bf (c)} $d_{1st}$ /Mpc: the projection distance of the nearest neighbor galaxy with 
  $m_r$ $<$ 17.7;  {\bf (d)} $N_{neigh}$: the total number of neighbor galaxies with $m_r$ $<$ 17.7, a line-of-sight velocity to the target galaxy within $\pm$500km s$^{-1}$, the projection distance within 1Mpc. The solid vertical lines represent the median value of different samples.}\label{vacia}
\end{figure*}

Figure \ref{vacia} shows the distribution of morphology, merging possibility and environmental parameters of Type Ia SN host galaxies and the control sample. The blue histogram represents the SN sample and the red represents the control sample. The vertical lines of different colors correspond to the median value of different distributions. In Figure \ref{vacia}(a), the median value of S{\`e}rsic index of Ia SN host galaxies is higher than the control sample with $n$ = 3.6 for SN sample and $n$ = 2.6 for controls, indicating that Type Ia SNe tends to explode in galaxies which are more bulge-dominated. Figure \ref{vacia}(b) shows that Type Ia hosts and their controls have very similar merging possibility ($P_{merg}$) distributions, indicating there is no obvious connection between merger/interaction and SN explosion. For environments, we find both $d_{1st}$ and $N_{neigh}$ shown in Figure \ref{vacia}(c) $\&$ (d) reflect similar distributions and median values, indicating a weak/no correlation between triggering of Type Ia SNe and the environments of their host galaxies.

\begin{figure*}
  \centering
  \includegraphics[width=0.9\textwidth]{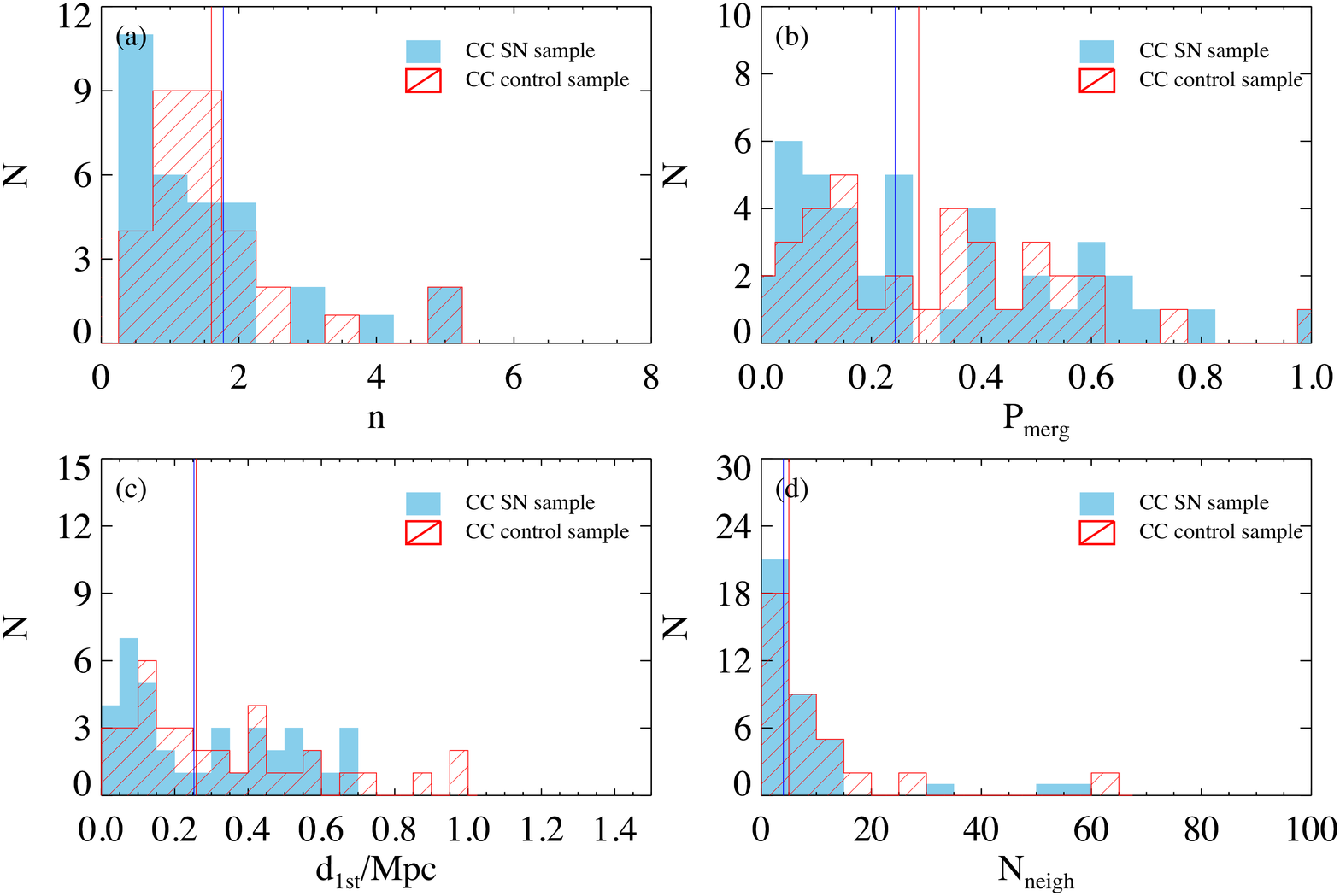}
  \caption{Similar to the Figure \ref{vacia}, but for Type CC SN host galaxies and their control sample.}\label{vaccc}
\end{figure*}

Similar to Figure \ref{vacia}, Figure \ref{vaccc} compared the Type CC SN host galaxies and their control sample, in which blue histograms represent Type CC hosts and the red represent the control galaxies, the two vertical lines of blue and red colors correspond to the median value of SN hosts and controls, respectively. Again, no difference is shown in Figure \ref{vaccc} indicating that there is no preference of SN explosion on the morphology, interaction and surrounding environment of their host galaxies.

\subsection{The Local Properties of SN Host Galaxies}

To further explore the local properties of the position where SNe explode, and to understand the speciality of the local environment, we select four regions in a galaxy for the comparison of galaxy properties: (i) the SN location: we select spaxels within $1^{\prime \prime}$ radius from MaNGA data around the SN location, taking the median value of the physical parameters as the local properties of SN regions; (ii) the galaxy center: we select the spaxels within $1^{\prime \prime}$ of galaxy center, taking the median parameters as the local properties of galaxy center region; (iii) the symmetric region of SN location relative to the center of the galaxy: we select the spaxels with $1^{\prime \prime}$ of the symmetric region of SN location relative to the center of the galaxy, taking the median parameters as the local properties; (iv) the annulus area with same effective radius as the SN location:   we select the spaxels at the annulus with the same effective radius as the SN position with a  width $\pm$0.05R$_e$, and take the median value as the properties of the annulus.

\begin{figure*}
  \centering
  \includegraphics[width=1\textwidth]{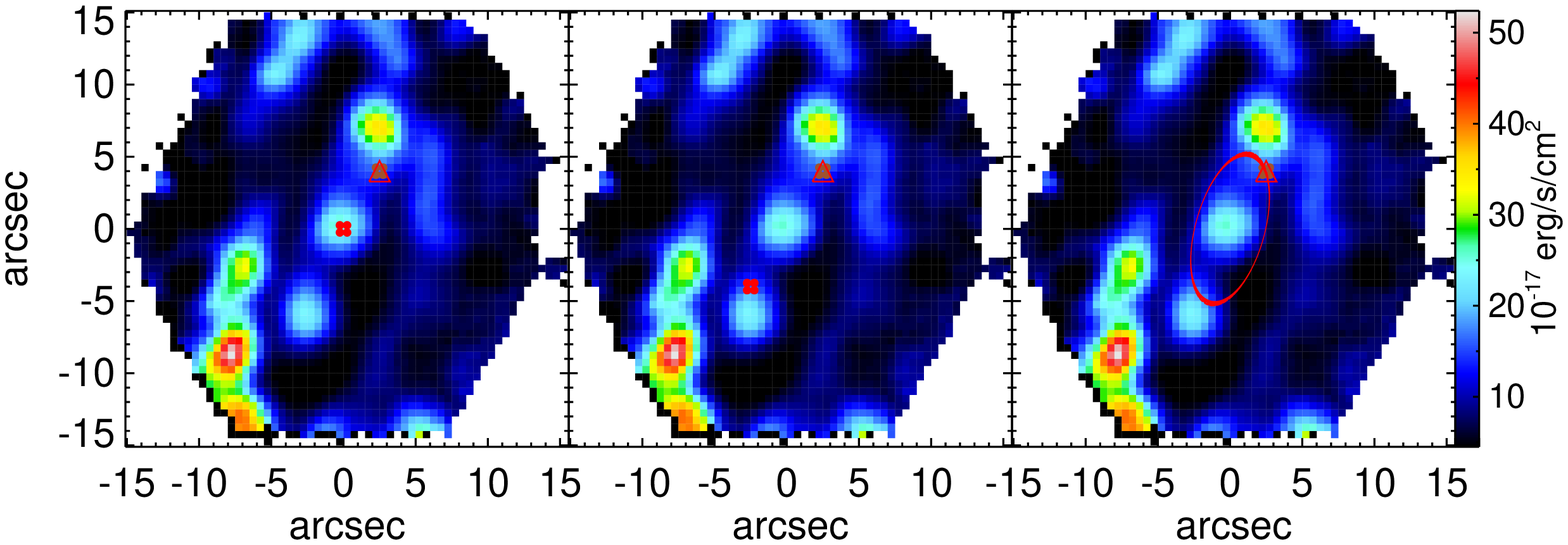}
  \caption{An example of H$\alpha$ flux map in MaNGA survey (MaNGAID: 1-115365). We mark the SN explosion location (red open triangle in all panels) and spaxels selected as SN explosion location (brown dots in all panels), the galaxy central region (red dots in left panel), the symmetric position of SN explosion region relative to the center of the galaxy (red dots in middle panel) and the annulus area with same effective radius as the SN explosion location (red ellipse in right panel).}\label{paint}
\end{figure*}

\begin{figure*}
  \centering
  \includegraphics[width=0.9\textwidth]{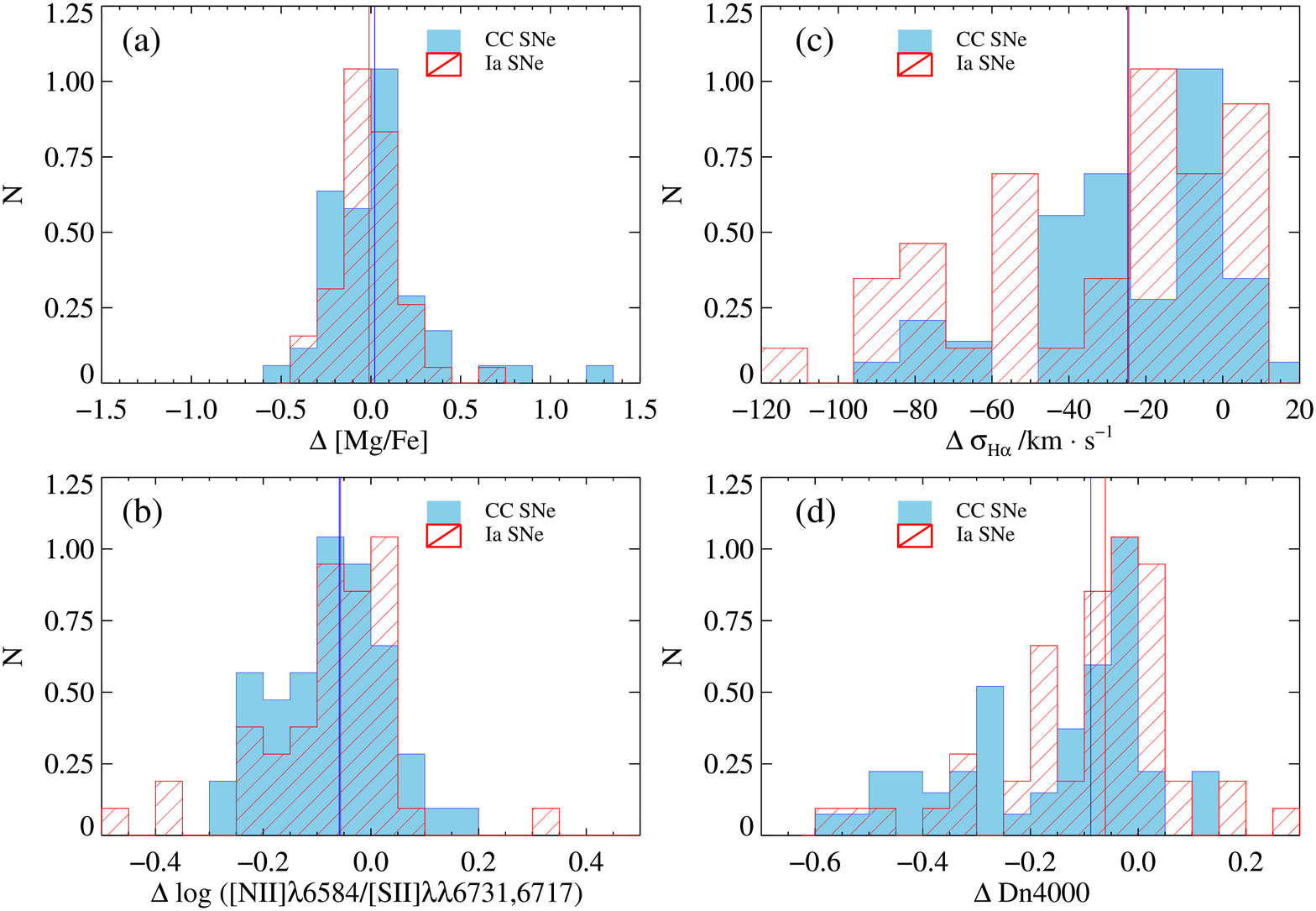}
  \caption{The distribution of the difference in [Mg/Fe] (a), \nii/\sii\  (b), $\sigma_{{\rm H\alpha}}$ (c), D$n$4000 (d) between SN position and the galaxy center. The red shaded histogram are for Type Ia SNe and blue for Type CC SNe. The vertical lines of different colors correspond to the median value of each distribution.}\label{centre}
\end{figure*}

Figure \ref{paint} is an example of H$\alpha$ flux map of SN host (MaNGAID: 1-115365). The SN location is marked as red triangles in each panel, the four brown dots are selected spaxels of the SN position. In the left panel, the four red dots in the middle represent the spaxels selected as galaxy center. At the bottom-left of the middle panel, the four red dots represent the data selected as symmetric position of SN explosion region relative to the center of the host galaxy. The red ellipse shown in right panel represents the data selected as the annulus area with same effective radius as the SN explosion location. The SN exploded in this galaxy is SN2007cn, which is classified as Type Ia.

We chose four parameters for comparison in this section: (i) we use [Mg/Fe] to roughly represent the star formation history. The magnesium element is an $\alpha$-process element, and the synthesis of the $\alpha$ element mainly comes from the explosion of Type CC SNe whose progenitors have a relatively short lifetime (about 1$\sim$10 Myr); Meanwhile most of the iron is formed from the explosion of Type Ia SNe whose progenitors are low mass star with a long lifetime. Due to the different time scale of their formation, the ratio [Mg/Fe] is a good tracer of the star formation history \citep{tho10}. The definition of [Mg/Fe] is:

\begin{equation}\label{mgfe}
\centering
\rm{[Mg/Fe]=Mgb/(0.5\times Fe\lambda5270+0.5\times Fe\lambda5335)}
\end{equation}
where the parameter Mgb, Fe$\lambda$5270 and Fe$\lambda$5335 are Lick indexes \citep{wor94} included in MaNGA DAP; (ii) we use \nii/\sii\  \citep{kew02,dop13} to characterize the gas-phase metallicity. We do not use metallicity indicators such as R$_{23}$ in this work since 
most of the spaxels in SN (especially type Ia) host galaxies are not star forming, the traditional gas-phase metallicity indicator becomes invalid. But \nii/\sii\ could overcome this problem in that nitrogen is a secondary $\alpha$-process element, and sulfur is a primary nuclear synthesis element. Although \nii/\sii\ is an approximation of N/O, at metallicity 12+log (O/H) $>$ 8.0, it has a good correlation with the O/H abundance \citep{dop16, kas16}. Since \nii\ and \sii\ have similar wavelengths, the reddening effects could also be neglected; (iii) we use ionized gas velocity 
dispersion $\sigma_{\rm H\alpha}$ to explore the influence of SN explosions on gas kinematics; (iv) D$n$4000 is a good indicator of stellar population age. It is defined by \citet{bal99} as the ratio of average flux density between 3850\r{A}$\sim$3950\r{A} and 4000\r{A}$\sim$4100\r{A}. 

Figure \ref{centre} shows distributions of the difference in [Mg/Fe], \nii/\sii\/, $\sigma_{\rm H\alpha}$, D$n$4000 between the SN exploding location and the galaxy center. The red shaded histograms are for Type Ia SNe and blue for Type CC SNe. A positive difference means the parameter at SN exploding location has higher value than that of galaxy center. The vertical lines of different colors correspond to the median value of each distribution. 
Figure \ref{centre}(a) shows the difference in [Mg/Fe]. The median value of both distributions are approaching 0 ($-0.01$ for Type Ia SNe and 0.02 for Type CC SNe), implying that there is no significant difference in star formation history between the SN position and the galaxy center. While in Figure \ref{centre}(b), we find that both Type Ia and Type CC SNe show obvious metallicity differences between the SN position and the galaxy center. The metallicity at SN position is lower than the galaxy center by $\sim$0.1 dex. Considering that the metallicity gradient of local star forming galaxies is typically $\sim$0.1 dex/R$_e$ \citep{sch20}, and the median value of SN position relative to the galaxy center is 0.8$\sim$0.9R$_e$ as shown in Figure \ref{eff}, this metallicity difference observed in Figure \ref{centre}(b) can be explained as a natural result of galaxy metallicity gradient.
Figure \ref{centre}(c) shows the difference in gas velocity dispersion. The median values of the differences in the Type Ia and Type CC SNe are $-18$km s$^{-1}$ and $-16$km s$^{-1}$, indicating that the gas velocity dispersion at the SN position is smaller than that of the galaxy center. This difference is mainly due to the larger contribution of the bulge at the galaxy center. The SN position locates mainly at the outer regions of galaxies where the contribution of the galaxy disk increases, and therefore has a lower velocity dispersion than the central area. 
Figure \ref{centre}(d) is the difference in D$n$4000. A negative value in $\Delta$D$n$4000 indicates that the stellar populations at SN position is younger than the galaxy center. SN position (with lower D$n$4000) is younger than the galaxy center. This is also a direct result of the stellar population gradient in a normal galaxy with the inside-out growth model \citep{fal80,mo98,som08}. In brief, the comparison of the parameters in the SN explosion position and the galaxy center shows that the observed differences can be naturally explained by the radial gradient of the parameters. It does not show either the special features of the SN position or the influence of a single SN explosion on the local physical properties of the host galaxy.

\begin{figure*}
  \centering
  \includegraphics[width=0.9\textwidth]{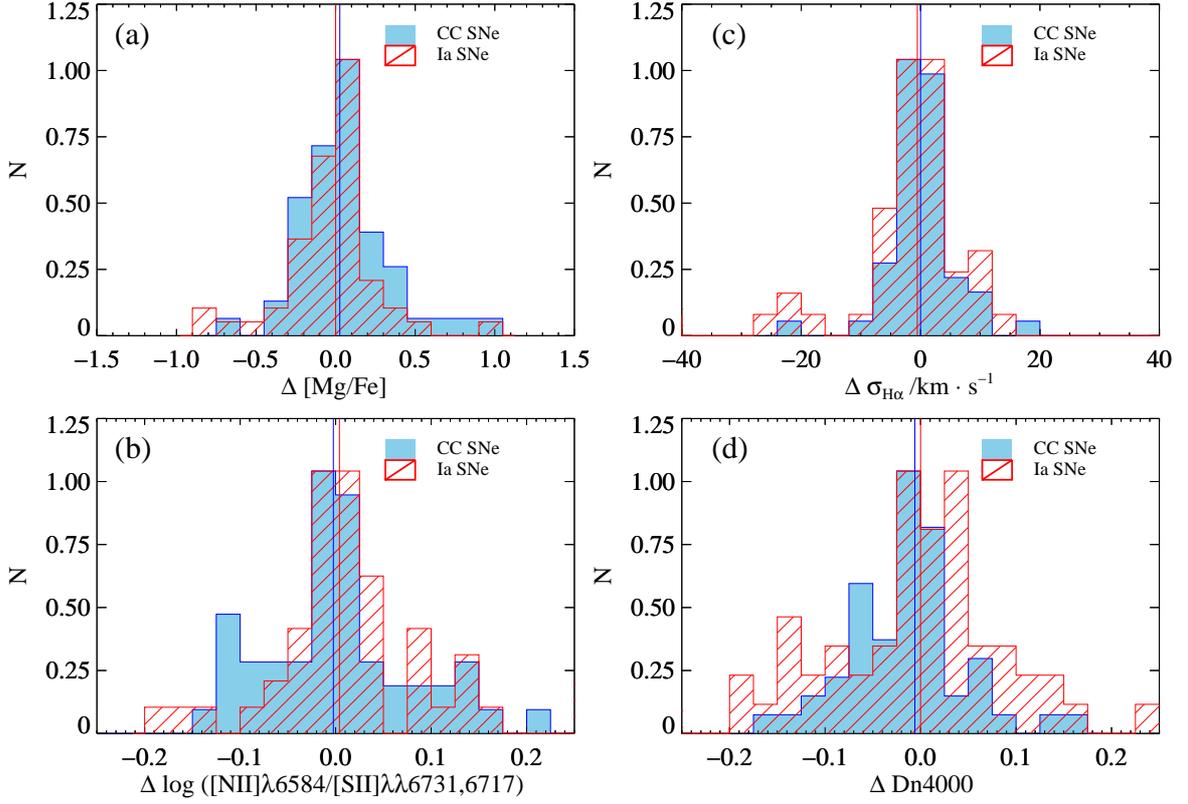}
  \caption{Similar to the Figure \ref{centre}. But for the differences in [Mg/Fe], \nii/\sii\/, $\sigma_{\rm H\alpha}$, D$n$4000 between the SN location and the symmetric region relative to the galaxy center.}\label{sym}
\end{figure*}

\begin{figure*}
  \centering
  \includegraphics[width=0.9\textwidth]{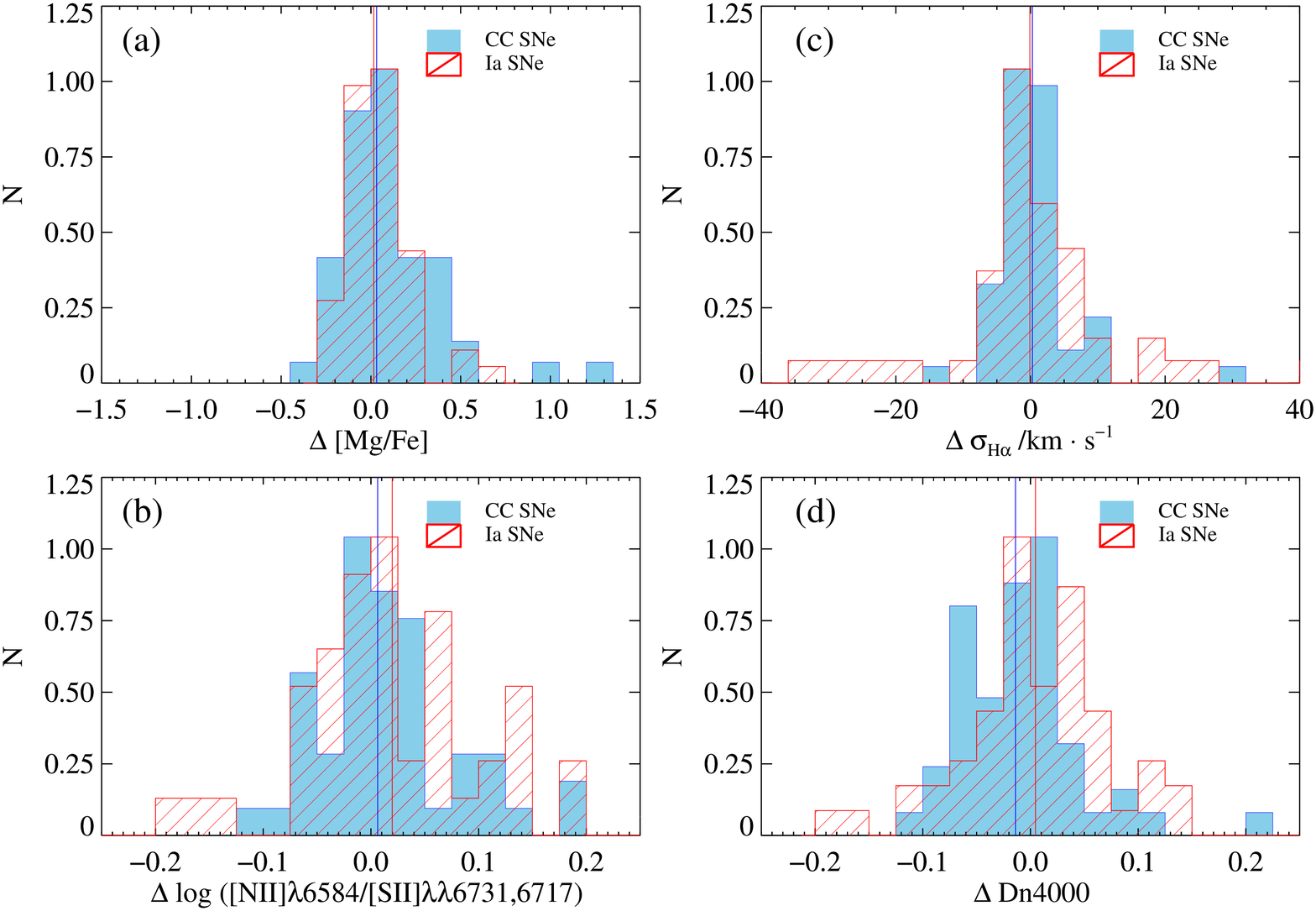}
  \caption{Similar to the Figure \ref{centre}. But for the differences in [Mg/Fe], \nii/\sii\/, $\sigma_{\rm H\alpha}$, D$n$4000 between the SN location and the annulus region with the same effective radius as SN position.}\label{ring}
\end{figure*}

In order to eliminate the influence of radial gradient, we compare the difference in the four parameters between SN location and its symmetrical region relative to the galaxy center. Statistically, the symmetrical region can be treated as a progenitor of SN location. The results are shown in Figure \ref{sym}. Similar to Figure \ref{centre}, The red shaded histograms are for Type Ia SNe and blue for Type CC SNe. The two vertical lines in each panel show the median value of the distributions with the same color. It is clear that the median values are all close to 0,
indicates that there is no significant difference between SN location and the symmetrical region. Considering that the symmetrical region 
could represent the state of the SN location before explosion, this result suggests: (i) the properties of the SN location do not show any particularities; (ii) the impact of a single SN explosion is not strong enough to be observed at the explosion region, or (iii) the spatial resolution is not higher enough to resolve the region that a single SN can affect. The typical spatial resolution of MaNGA is $\sim$1.2 kpc considering a $2^{\prime \prime}$ fiber size at a median redshift of $z\sim$0.03. We are looking forward to higher spatial resolution IFU observations for further research.

Currently, the definition of symmetrical region covers only 3$\sim$4 spaxels, for which the physical properties could be influenced by substructures. Thus we further compare the average properties of SN location and elliptical annulus with the same effective radius as SN position. Similar to Figure \ref{centre}, Figure \ref{ring} shows the distributions of difference between the median value of annulus region and the SN position in [Mg/Fe], $\sigma_{\rm H\alpha}$, \nii/\sii\ and D$n$4000. We find the difference of these parameters also have a value nearly zero, which again points to the result that the properties of SN location is similar to regions at the same effective radii. This result is consistent with \citet{galb14,galb16} which found that the SFR and metallicity of the SN location are almost the same as that of the global properties of the galaxy based on the CALIFA survey.

\section{Discussion}

\subsection{Sample Size}

\begin{figure*}
  \centering
  \includegraphics[width=0.7\textwidth]{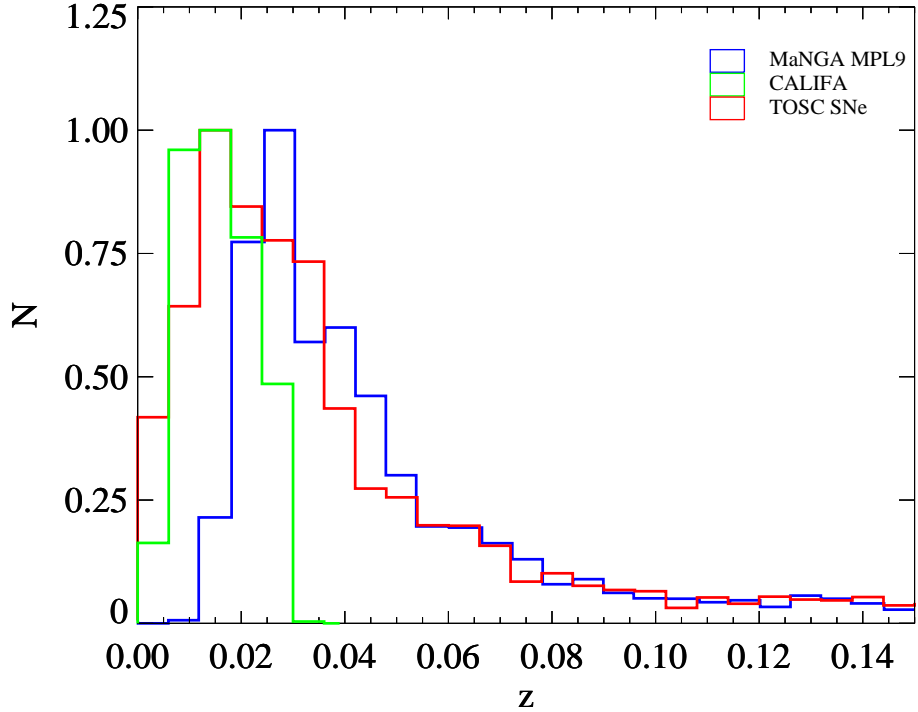}
  \caption{Redshift distribution of MaNGA MPL9 sample (blue), CALIFA sample (green) and TOSC SN sample (red). The peaks of all the distributions are set to 1.}\label{red}
\end{figure*}

In this work, we find 132 SNe with classification within the MaNGA bundle size from 8000 unique MaNGA galaxies, corresponds to $\sim$2$\%$ SN fraction. However, \cite{galb16} found a SN fraction of  $\sim$14$\%$ (= 132/939) from CALIFA survey. To understand the large difference in SN fraction, we explore the redshift distribution of MaNGA MPL9 (blue), CALIFA (green), and TOSC (red) samples in Figure \ref{red}.  We find that the redshift of the CALIFA sample is mainly distributed at  a redshift smaller than 0.03, and peaks $\sim$0.015. The redshift coverage of MaNGA is 0.01$\sim$0.15 and peaks at $\sim$0.03 with very few galaxies at $z<$0.015. There are 10,633 SNe in TOSC which has a clear SN type classification and a specific host galaxy. The peak of their host galaxy redshift distribution is around 0.015, similar to CALIFA. The large difference in SN fraction can be easily understood by the different redshift distributions: galaxy sample from CALIFA survey has similar redshift distribution as the SN host galaxies, while MaNGA galaxies are located at higher redshifts. The lower SN fraction is due to this redshift difference between MaNGA galaxy sample and SN hosts. This result is totally consistent with \cite{zho19} who found a SN fraction of 1$\%$ (= 11/1390) in early MaNGA data release.

\subsection{The Stellar Mass and Morphology of SN Host Galaxies}

In this work, we study the distribution of supernova host galaxies on SFR vs. stellar mass relation, and find that the Type CC SN host galaxies are distributed in the star-forming main sequence, while Type Ia SNe are distributed in both star-forming galaxies and quiescent galaxies. This is consistent with the result of \citet{sha14} who found CC SNe mostly lie in star-forming galaxies and Ia SNe lie in star-forming, AGN, weak emission-line galaxies as well as absorption-line galaxies. Meanwhile, the increasing of the Type Ia/CC number ratio with stellar mass in this work suggests that Type Ia SNe tend to explode in galaxies with higher masses, which is in line with the result of \citet{zho20}.

We compare the global properties of SN host galaxies and control sample, finding that there is no obvious difference in galaxy interactions and surrounding environments. The S{\`e}rsic index of Type Ia SN host galaxies is slightly higher than that of their control sample, indicating that Type Ia supernovae are more likely to explode in galaxies that are more bulge-dominated. However, \citet{man05} found that the SN explosion rate is $\sim$20 times higher in spiral galaxies than that of elliptical galaxies. This result, at first sight, is in conflict with our result. Actually, this difference is due to that we are comparing the Type Ia hosts and their control galaxies with similar stellar mass and SFR, while \citet{man05} compared SN explosion rate in spiral and elliptical galaxies. In Figure \ref{contour1}, we can find that there are 48 Type Ia SNe in star forming main sequence, and 19 Ia SNe located in quiescent sequence below the red dashed line, indicating the Type Ia SN explosion rate is higher for star forming main sequence galaxies, which is in consistent with \citet{man05} qualitatively.

\section{Conclusion}

In this paper, we crossmatch galaxies from MaNGA MPL9 to SNe from TOSC, obtaining a total of 132 SNe with classifications within MaNGA bundle. These 132 SNe can be classified into 67 Type Ia and 65 Type CC, where Type CC includes Type Ib, Ic, and Type II SNe. Type Ia SNe are distributed in both star-forming galaxies and quiescent galaxies, while Type CC SNe are all distributed along the star-forming main sequence. As the stellar mass increases, the Type Ia/CC number ratio also increases. Our conclusions are summarized as follows:

\begin{enumerate}
  \item About the global properties of SN host galaxies, we find that there is no obvious difference in the interaction possibilities and environments between Type Ia SN hosts and a control sample of galaxies with similar stellar mass and SFR distributions, except that Type Ia SNe tend to appear in galaxies which are more bulge-dominated than their controls. For Type CC SNe, there is no difference between their host and the control galaxies in galaxy morphology, interaction possibility as well as environments. These results indicate the SN explosion rate is not depending on galaxy interactions and environments.
  \item About the local properties, we find the SN locations with smaller velocity dispersion, lower metallicity, and younger stellar population than galaxy centers. This is a result of radial gradients of all these parameters. The comparison of physical properties ([Mg/Fe], \nii/\sii\/, $\sigma_{\rm H\alpha}$ and D$n$4000) between SN locations with the symmetric regions relative to galaxy centers and elliptical annulus with similar effective radii as SN location finds no obvious difference, indicating  either there is no particularity in the properties at SN location or the scale that a single SN explosion can influence is much smaller than kpc, the typical resolution of MaNGA.
  \end{enumerate}

\normalem
\begin{acknowledgements}
We thank Yongyun Chen, Jianhang Chen and Songlin Li for very constructive comments in our work. 
Y. C acknowledges support from the National Key R\&D Program of China (No. 2017YFA0402700), the National Natural Science Foundation of China (NSFC grants 11573013, 11733002, 11922302).

Funding for the Sloan Digital Sky Survey IV has been provided by the Alfred P.
Sloan Foundation, the U.S. Department of Energy Office of Science, and the Participating Institutions.
SDSS-IV acknowledges support and resources from the Center for High-Performance Computing at
the University of Utah. The SDSS web site is www.sdss.org.

SDSS-IV is managed by the Astrophysical Research Consortium for the Participating Institutions of the SDSS Collaboration 
including the Brazilian Participation Group, the Carnegie Institution for Science, Carnegie Mellon University, the Chilean Participation 
Group, the French Participation Group, Harvard-Smithsonian Center for Astrophysics, Instituto de Astrof\'isica de Canarias, The Johns 
Hopkins University, Kavli Institute for the Physics and Mathematics of the Universe (IPMU) / University of Tokyo, Lawrence Berkeley 
National Laboratory, Leibniz Institut f\"ur Astrophysik Potsdam (AIP), Max-Planck-Institut f\"ur Astronomie (MPIA Heidelberg), 
Max-Planck-Institut f\"ur Astrophysik (MPA Garching), Max-Planck-Institut f\"ur Extraterrestrische Physik (MPE), National 
Astronomical Observatories of China, New Mexico State University, New York University, University of Notre Dame, 
Observat\'ario Nacional / MCTI, The Ohio State University, Pennsylvania State University, Shanghai Astronomical Observatory, 
United Kingdom Participation Group, Universidad Nacional Aut\'onoma de M\'exico, University of Arizona, University of Colorado Boulder, 
University of Oxford, University of Portsmouth, University of Utah, University of Virginia, University of Washington, University of Wisconsin, 
Vanderbilt University, and Yale University.

\end{acknowledgements}
  
\bibliographystyle{raa}
\bibliography{ms2021-0220}

\clearpage
\onecolumn
\appendix
\section{ }

{
\small
\begin{center}
  \begin{longtable}{cccccc}
  \caption{Basic information for SN sample. The plate-ifudsgn column refers to the SN host galaxy observed in MaNGA.}\label{table:sne}\\
    \toprule
    \textbf{SN Name} & \textbf{plate-ifudsgn} & \textbf{SN R.A.} & \textbf{SN Dec.} & \textbf{SN Type} & \textbf{redshift} \\
    \midrule
    \endfirsthead
    
    \toprule
    \textbf{SN Name} & \textbf{plate-ifudsgn} & \textbf{SN R.A.} & \textbf{SN Dec.} & \textbf{SN Type} & \textbf{redshift} \\
    \midrule
    \endhead
    
    \bottomrule
    \endfoot
                          SN2010ee  &   7495-12702    &  205.074951  & 26.353306  &  II  &  0.028401 \\
                     SN2017frb  &   7815-3702     &  317.903473  & 11.497245  &  Ia  &  0.029382 \\
                      SN2007cn  &   7979-12701    &  333.482452  & 13.756639  &  Ia  &  0.025379 \\
                     SN2017jcu  &   8083-12704    &   50.695415  &  0.148083  &  II  &  0.022831 \\
                       SN2007R  &   8138-12704    &  116.656372  & 44.789471  &  Ia  &  0.030805 \\
                     SN2019smv  &   8258-3701     &  165.569458  & 44.334660  &  II  &  0.024607 \\
                      SN2012hj  &   8257-12704    &  166.829956  & 46.379501  &  Ia  &  0.024578 \\
                     SN2018hfc  &   8257-12701    &  165.494095  & 45.227539  &  II  &  0.019996 \\
                     SN2018ats  &   8453-3701     &  153.234543  & 46.418091  &  Ia  &  0.038213 \\
                      SN2012al  &   8453-12702    &  151.548370  & 47.294613  &  II  &  0.038107 \\
                       SN2010D  &   8452-12705    &  157.934494  & 46.668056  &  II  &  0.024944 \\
                     SN2018aex  &   8448-6102     &  165.189087  & 22.287500  &  II  &  0.022903 \\
                      PTF11bui  &   8465-12705    &  198.235001  & 47.453472  &  Ia  &  0.028138 \\
                     SN2019cvz  &   8484-12705    &  247.725021  & 46.588387  &  II  &  0.018346 \\
                     SN2019gqd  &   8312-12703    &  247.208862  & 39.834347  &  Ic  &  0.035850 \\
                    SN2020abla  &   8312-3704     &  247.164597  & 39.521030  &  Ia  &  0.025177 \\
                      SN2015cq  &   8550-9101     &  247.410828  & 40.236111  &  II  &  0.028330 \\
                      SN2005bp  &   7957-12704    &  257.430084  & 36.418915  &  II  &  0.027672 \\
                       PTF09sh  &   8314-6101     &  243.492004  & 39.532806  &  II  &  0.037690 \\
                      SN2008ev  &   8314-12701    &  240.758713  & 39.645611  &  II  &  0.030927 \\
                      SN2000cs  &   8604-12701    &  245.884338  & 39.124748  &  II  &  0.035039 \\
                      SN1999cb  &   8604-9102     &  246.451706  & 40.342335  &  Ia  &  0.029039 \\
                     SN2018aej  &   8315-12704    &  236.095871  & 39.558094  &  Ia  &  0.047934 \\
                       PS15aot  &   8603-6101     &  247.159836  & 39.548668  &  Ia  &  0.031176 \\
                     SN2020mnv  &   8601-9101     &  250.182632  & 40.414482  &  Ia  &  0.025886 \\
                      SN2011cc  &   8588-6101     &  248.455994  & 39.263527  &  II  &  0.031761 \\
                   ASASSN-15ns  &   8588-6102     &  250.116745  & 39.320221  &  Ia  &  0.030079 \\
                     SN2019eho  &   8615-9102     &  321.472076  &  0.414669  &  Ia  &  0.031740 \\
                     SN2017fel  &   8616-12702    &  322.305756  & -0.294703  &  Ia  &  0.030537 \\
                      PTF12izc  &   8655-3702     &  355.893219  &  0.568667  &  Ia  &  0.082650 \\
                      SN2015co  &   7960-12705    &  258.650543  & 30.735666  &  II  &  0.029594 \\
                     SN2020jck  &   7968-12701    &  322.948822  & -1.039164  &  Ia  &  0.051547 \\
                      SN2010dl  &   7968-9102     &  323.754028  & -0.513278  &  Ia  &  0.030156 \\
                       SN2004I  &   8078-12701    &   40.880123  &  0.308528  &  II  &  0.026670 \\
                      PTF12iiq  &   8078-6102     &   42.532333  & -0.265111  &  Ia  &  0.029084 \\
                     SN2019omi  &   8086-9102     &   57.249107  &  0.926919  &  Ia  &  0.035788 \\
                      LSQ12fnt  &   8155-12701    &   53.172501  & -1.186389  &  Ic  &  0.030451 \\
                     SN2017ixz  &   8147-12701    &  116.762627  & 26.773825  &  II  &  0.023592 \\
                     SN2017ckx  &   8145-1902     &  117.045876  & 28.230289  &  Ia  &  0.027158 \\
                     SN2017cfq  &   8149-9102     &  120.980026  & 26.520203  &  II  &  0.021748 \\
                     SN2020cvy  &   8149-3704     &  120.205307  & 27.498581  &  II  &  0.017345 \\
                     SN2019abu  &   8084-12702    &   50.536613  & -0.840061  &  II  &  0.036459 \\
                     SN2018ddh  &   8262-6104     &  184.683502  & 44.781975  &  Ia  &  0.038341 \\
                     SN2021drc  &   8450-6103     &  171.954544  & 21.386717  &  Ia  &  0.040623 \\
                     SN2021efd  &   8450-9101     &  171.103043  & 23.649530  &  Ib  &  0.027756 \\
                   ASASSN-15pn  &   8158-1901     &   60.859665  & -5.492028  &  Ia  &  0.038357 \\
                      LSQ12hxg  &   8158-9102     &   62.947582  & -5.919028  &  II  &  0.038794 \\
                     AT2017boa  &   8547-12701    &  217.631790  & 52.704327  &  Ia  &  0.044881 \\
                      SN2002aw  &   8978-6102     &  249.371078  & 40.880638  &  Ia  &  0.026385 \\
                     SN2020dow  &   8987-9102     &  137.983322  & 27.900387  &  Ia  &  0.047433 \\
                     SN2018btb  &   8945-9102     &  173.616257  & 46.362530  &  Ia  &  0.033842 \\
                    SN2020acfp  &   8945-12701    &  171.894165  & 47.379429  &  Ia  &  0.032730 \\
                      SN1998cs  &   9029-12701    &  247.663422  & 41.215111  &  Ia  &  0.032463 \\
                      SN2002cg  &   9029-12702    &  247.251999  & 41.283443  &  Ic  &  0.031887 \\
                      SN2004ct  &   9041-9102     &  235.940628  & 28.416529  &  Ia  &  0.032751 \\
                     SN2017dgs  &   9041-12702    &  236.389847  & 30.145430  &  II  &  0.031628 \\
            SDSS1684-53239-484  &   9027-1901     &  245.686615  & 32.659184  &  II  &  0.040998 \\
                      SN2012fj  &   9027-12701    &  243.936920  & 31.963249  &  II  &  0.031547 \\
                      SN2002ci  &   9027-9101     &  243.908127  & 31.321501  &  Ia  &  0.022198 \\
                      SN2005bk  &   9036-9102     &  240.570999  & 42.915359  &  Ic  &  0.024435 \\
         PSN-J12565170+2629167  &   8951-12705    &  194.215424  & 26.487972  &  Ib  &  0.025364 \\
                     SN2017cxz  &   8625-9101     &  259.832428  & 57.898834  &  Ib  &  0.028949 \\
              SDSS-II-SN-20130  &   7964-12704    &  316.558014  &  0.098017  &  II  &  0.050593 \\
                     SN2019abp  &   9050-6102     &  245.860565  & 22.486141  &  II  &  0.037521 \\
                      SN2007gl  &   8080-12705    &   47.888500  & -0.746412  &  Ib  &  0.028230 \\
                     SN2018jfp  &   8080-12703    &   49.484486  & -0.169706  &  II  &  0.022797 \\
                       SN2021T  &   8657-9101     &    9.829108  & -0.455769  &  Ia  &  0.045407 \\
              SDSS-II-SN-12882  &   8154-9102     &   45.958199  & -0.203996  &  II  &  0.027592 \\
                     SN2020ksa  &   8993-12701    &  164.866455  & 46.124577  &  Ib  &  0.022165 \\
                     SN2016acq  &   9509-12703    &  124.102455  & 25.993111  &  II  &  0.045242 \\
                     SN2016gkm  &   9501-12705    &  130.721100  & 25.070923  &  Ib  &  0.017257 \\
                      SN2003an  &   8984-9101     &  201.973129  & 28.508112  &  Ia  &  0.036994 \\
                     SN2020fcx  &   8311-12703    &  205.041748  & 23.341530  &  II  &  0.031926 \\
                       SN2007O  &   9033-12705    &  224.021576  & 45.404694  &  Ia  &  0.036187 \\
                       PTF11go  &   8996-6102     &  173.001007  & 53.710583  &  II  &  0.026810 \\
                     SN2021bqv  &   9038-6101     &  238.758133  & 41.578018  &  Ia  &  0.033563 \\
                      PTF11cao  &   9048-12704    &  244.699829  & 25.187916  &  Ia  &  0.039508 \\
                     SN2020uoo  &   9095-6102     &  243.442657  & 22.919676  &  Ia  &  0.031916 \\
                       SN2002G  &   8323-6101     &  196.980255  & 34.085140  &  Ia  &  0.033650 \\
                     SN2020ofw  &   9863-12701    &  195.139450  & 27.504499  &  II  &  0.018608 \\
                       SN2009L  &   9876-12703    &  194.700424  & 27.673779  &  Ia  &  0.027962 \\
                     SN2021apg  &   9881-12704    &  205.330185  & 24.495543  &  II  &  0.026856 \\
                     SN2018lev  &   9871-12702    &  228.284042  & 41.263699  &  II  &  0.029064 \\
                      PTF12ewk  &   9869-12702    &  247.241119  & 39.319721  &  Ia  &  0.033924 \\
                      SN2006cq  &   8322-12705    &  201.104584  & 30.956306  &  Ia  &  0.048502 \\
                     SN2020kte  &   9186-6102     &  259.121674  & 29.445555  &  Ia  &  0.030187 \\
                     SN2018emi  &   9893-9102     &  256.651367  & 24.545319  &  Ia  &  0.038230 \\
                      PTF13dfa  &   8097-6104     &   27.377167  & 13.992556  &  Ia  &  0.068000 \\
                     SN2019vin  &   9189-12705    &   52.705372  & -7.046347  &  Ia  &  0.038909 \\
                     SN2019smi  &   8986-6101     &  119.868462  & 26.889795  &  Ic  &  0.026470 \\
                     SN2020umy  &   8986-6104     &  119.902893  & 28.863256  &  II  &  0.045626 \\
                      SN2006te  &   10213-12705   &  122.929123  & 41.554668  &  Ia  &  0.031548 \\
                      SN2007kd  &   8152-12701    &  141.491714  & 34.633141  &  Ia  &  0.024224 \\
                      SN2012cq  &   8245-12701    &  137.022751  & 20.503471  &  II  &  0.025612 \\
                     SN2019dfa  &   8438-12702    &  149.749191  & 17.818794  &  Ia  &  0.025439 \\
                     SN2016aak  &   8248-12705    &  138.554214  & 16.741722  &  Ia  &  0.027855 \\
                     SN2020hgg  &   8988-9102     &  187.235001  & 40.847279  &  Ia  &  0.064948 \\
                      SN2020fc  &   10519-12704   &  154.897339  &  6.328467  &  II  &  0.028476 \\
                     SN2019vsi  &   9882-12701    &  206.479034  & 23.061930  &  Ib  &  0.027886 \\
                     SN2019hjl  &   10503-6104    &  160.622864  &  5.083230  &  II  &  0.028108 \\
                       SN2012O  &   9880-12701    &  195.166626  & 27.923445  &  Ia  &  0.025087 \\
                      SN2011ki  &   9040-12701    &  242.626923  & 27.499777  &  II  &  0.031851 \\
                     SN2021dxo  &   8260-6101     &  182.411789  & 42.008957  &  II  &  0.022881 \\
                      SN2018ds  &   8441-12701    &  222.223206  & 38.767693  &  Ia  &  0.031662 \\
                     SN2019lpd  &   9030-9102     &  243.570999  & 30.999445  &  Ia  &  0.061721 \\
                      SN1999df  &   8593-6104     &  227.235504  & 52.448082  &  II  &  0.037914 \\
                      SN2001it  &   8593-12704    &  226.539169  & 53.411556  &  II  &  0.034190 \\
                      SN2007ac  &   11945-12701   &  251.759827  & 40.146557  &  II  &  0.030193 \\
                     SN2020ytg  &   9191-6101     &  312.041656  & -1.211139  &  II  &  0.056335 \\
                      SN2013ag  &   11871-6102    &  192.895920  & 26.629278  &  Ia  &  0.021295 \\
                      SN2012ee  &   9187-9101     &  311.313660  & -5.622806  &  Ia  &  0.027326 \\
                      LSQ12btw  &   10518-3703    &  152.620087  &  5.536800  &  Ib  &  0.057650 \\
                      SN2006bz  &   11006-12704   &  195.180756  & 27.961611  &  Ia  &  0.028093 \\
                     SN2019gmh  &   11941-12705   &  247.763168  & 41.153965  &  II  &  0.030194 \\
                      SN1992ap  &   11941-12704   &  247.613876  & 41.487804  &  Ia  &  0.029864 \\
                     iPTF13bld  &   11942-12701   &  246.227448  & 41.049789  &  II  &  0.033074 \\
                      SN2009fv  &   11942-12705   &  247.434250  & 40.811611  &  Ia  &  0.029338 \\
                      SN2008cw  &   11942-6102    &  248.159454  & 41.459221  &  II  &  0.032433 \\
                      SN2014bo  &   11942-12703   &  246.942291  & 41.739918  &  Ia  &  0.133471 \\
                       SN2008Z  &   11751-12703   &  145.813538  & 36.284416  &  Ia  &  0.020621 \\
                     SN2020gac  &   11755-3703    &  187.442627  & 42.732700  &  II  &  0.038634 \\
                      PTF11bnf  &   8563-12701    &  241.186752  & 49.457832  &  Ia  &  0.024164 \\
                       SN1940C  &   8589-12705    &  226.729172  & 56.508888  &  II  &  0.029552 \\
                     SN2019qsc  &   8589-6104     &  226.512772  & 55.573055  &  Ia  &  0.029223 \\
                      PTF10bhu  &   11761-9101    &  193.868500  & 53.574638  &  Ic  &  0.035834 \\
               SNF20080322-000  &   11979-12704   &  252.089371  & 23.614750  &  Ia  &  0.030252 \\
                      SN2001dy  &   11984-6103    &  256.247589  & 23.168388  &  II  &  0.030081 \\
                      SN2010ed  &   11984-12704   &  257.350555  & 22.213362  &  Ia  &  0.048984 \\
                     SN2019gvw  &   11827-12702   &  211.810120  & 49.093830  &  Ia  &  0.071327 \\
                     SN2016acx  &   11014-1902    &  194.854996  & 27.740101  &  Ia  &  0.020117 \\
                     SN2018iun  &   11018-3703    &  197.128494  & 50.641460  &  II  &  0.029314 \\
                     SN2007F  &   11018-12704   &  195.812836  & 50.618805  &  Ia  &  0.023619 \\
\bottomrule
\end{longtable}
\end{center}
}
\end{document}